\documentclass[12pt]{article}

\input epsf
\def\beq{\begin{eqnarray}}
\def\eeq{\end{eqnarray}}

\def\mpl{M_{\rm Pl}}
\def\e{{\cal E}}

\def\lsim{\mathrel{\rlap{\lower3pt\hbox{\hskip0pt$\sim$}}
     \raise1pt\hbox{$<$}}}         
\def\gsim{\mathrel{\rlap{\lower4pt\hbox{\hskip1pt$\sim$}}
     \raise1pt\hbox{$>$}}}         

\def\ud{{\rm d}}
\def\uD{{\rm D}}

\usepackage{array}
\usepackage{amsmath}
\usepackage{amsfonts}
\usepackage[dvips]{graphicx}

\numberwithin{equation}{section}

\newcommand{\qket}[1]{\left|{#1}\right\rangle}

\begin{document}

\begin{flushright}
{ NYU-TH-05/10/11}
\end{flushright}
\vskip 0.9cm

\centerline{\Large \bf Quantum Cosmology of Classically Constrained 
Gravity}
\vspace{0.3in}
\vskip 0.7cm
\centerline{\large Gregory Gabadadze and Yanwen Shang}
\vskip 0.3cm
\centerline{\em Center for Cosmology and Particle Physics}
\centerline{\em Department of Physics, New York University, New York, 
NY, 10003, USA}

\vskip 1.9cm

\begin{abstract}

In hep-th/0506040 we discussed a classically constrained model 
of gravity. This theory contains known 
solutions of General Relativity (GR), and admits solutions 
that are absent in GR. Here we study cosmological implications of 
some of these new solutions. We show that a 
spatially-flat de Sitter universe can be  created  from ``nothing''.  
This universe has boundaries, and its total energy equals to zero. 
Although the probability to create such a universe
is exponentially suppressed, it favors initial conditions 
suitable for inflation. Then we discuss a finite-energy solution  
with a nonzero cosmological constant and zero space-time curvature.
There is no tunneling suppression to fluctuate into this state.
We show that for a positive cosmological constant
this state is unstable -- it can  rapidly transition  to a de 
Sitter universe  providing a new unsuppressed channel for inflation. 
For a negative cosmological constant 
the space-time flat solutions is stable.

\end{abstract}

\newpage

\section{Introduction}

In Ref. \cite {GShang} a classically constrained General Relativity
(CGR) was discussed. The gravitational part of the  Lagrangian density 
consists of the conventional Einstein-Hilbert (EH) 
term amended by a term that enforces a constraint
\beq
\mathcal{L}=-{\sqrt{-g}\over 2} (R+2\Lambda)
- \sqrt{-g}g^{\nu\mu}\partial_\nu\lambda_\mu
+\dots .
\label{Lagr}
\eeq
Here $\lambda_\mu$  is a non-dynamical 
Lagrange multiplier field,  and we introduced a 
cosmological constant $\Lambda$. In most of the applications 
discussed below, $\Lambda$ can be replaced by a ``slow roll'' 
inflationary potential $V(\phi)$, 
as $\Lambda \to V(\phi)$ (we put $\mpl=1$).

The Lagrangian (\ref {Lagr}) is a part of the action 
used in  path-integral quantization of 
GR\footnote{GR is not a renormalizable theory, however, 
it can be regarded as a low-energy
effective quantum field theory with a cutoff. For an 
exposition of this point of view, see Ref. \cite {Donoghue}.}.
The Lagrange multiplier term  usually enforces 
the gauge fixing condition.  For consistent quantization of small 
fluctuations this Lagrangian should be amended
by appropriate boundary conditions for the fluctuations, and 
by the Faddeev-Popov (FP) ghosts.
The main point of the  approach  of Ref. \cite {GShang}, which we 
follow  here, was  to allow for the boundary conditions on which the 
determinant  of the FP operator has a zero-mode. This would make 
the path integral ill-defined, unless the zero-mode  
is treated separately from the fluctuations.  
The zero-mode is regarded as a classical background solution, 
and the small fluctuations are then quantized about 
that background. In Ref. \cite {GShang} we considered  only the 
background  solutions on which the FP ghosts vanish, although 
they are present as quantum fluctuations.

The above approach, when it comes to classical solutions, 
reduces to the  following simple algorithm.
Considering (\ref {Lagr}) as a classically constrained theory. 
In this theory, Einstein's equations are modified due to the 
$\lambda_\mu$ field. The  modified equations 
could allow for new solutions \cite {GShang} that are absent in GR.   
To discuss those solutions we consider spaces with  boundaries where  the 
Gibbons-Hawking term \cite {GH} is implied and 
the following boundary conditions are imposed: 
$\delta g_{\mu\nu}|_{\rm boundary}=\delta \lambda_\mu|_{\rm boundary} =0$.
Then, the  equations of motion take the form:
\beq
G_{\mu\nu}+(\partial_\mu\lambda_\nu+\partial_\nu\lambda_\mu)-
g^{\sigma\tau}\partial_\sigma
\lambda_\tau g_{\mu\nu}&= &\Lambda g_{\mu\nu},
\label{eq1}\\
\partial_\mu(\sqrt{-g}g^{\mu\nu})&=&0.
\label{gauge}
\eeq
The above equations  can  admit solutions that are not present 
in GR. For instance, a theory with  $\Lambda \neq 0$ has a 
solution with zero space-time 
curvature \cite {GShang}:  $g_{\mu\nu}=\eta_{\mu\nu},~\partial_\mu 
\lambda_\nu  + \partial_\nu 
\lambda_\mu= -\Lambda \eta_{\mu\nu}$. The solution 
ends on  a fixed boundary where the value of $\lambda_\mu$,
which is defined up to a constant, is adjusted to be zero 
so that the space is geodesically complete.  
We will call this a new flat solution below.
On the other hand, putting  $\lambda_\mu=0$, 
the theory  yields a conventional (anti)de Sitter solution 
written in  a gauge (\ref {gauge}) \cite {GShang}. 
There are also other solutions, one of them being 
a zero-energy   spatially-flat de Sitter (dS) 
space with a boundary, that  we will discuss 
below in some detail.

The goal of the  present work is to study these solutions and their relevance  
to cosmology.  As a first example, we will look at 
a new possibility to create a universe 
into a state described by the {\it spatially-flat} dS solution. That 
quantum creation of a spatially-flat universe is possible if it 
has non-trivial topology,  was first found by Zel'dovich and Starobinsky
\cite {ZS}. In our case, the spatially-flat universe 
that is being created has trivial topology, but comes with a fixed 
boundary  on which the boundary conditions preserving
completeness of the space are imposed. We calculate  the probability of 
creation of  such a universe  out of ``nothing'', i.e., out of an 
initial state with no classical  space-time. 
Linde's \cite {Linde} and Vilenkin's 
(first reference in \cite {Vilenkin}) approaches  
give the same results in this case. We will find that the probabilistic  
arguments favor initial conditions needed for inflation,
as opposed to the conditions that would favor
universe sitting at the bottom of the potential. However, 
the probability itself is still exponentially small.
This is somewhat similar to the emergence  of a {\it closed} 
dS universe in a conventional approach \cite {Linde,Vilenkin}.

Then we turn to a new flat solution described above.
We study a process of producing a small region of primordial 
universe in a state of  a nonzero energy described by the new flat 
solution. As we will see, in the minisuperspace approximation,
there is no potential  barrier to be penetrated in order
to fluctuate into this state. Interestingly enough,  if 
$\Lambda$ is positive, this state is unstable -- 
it can either collapse or  with an almost equal probability, 
can  rapidly  transition into a spatially-flat dS  
universe with  $H^2 = \Lambda/3$.  The latter 
can be used to describe the required inflationary 
epoch.  The above sequence of events, represents a new channel 
for obtaining an inflationary region in a  primordial universe. 
The probability of these events  to take place is 
not suppressed by the exponential factors. In that regard,
the effect is similar to the one emphasized  
by Linde \cite {LindeFlat}, in the context of the 
solution of \cite {ZS}\footnote{We thank A. Vilenkin for 
bringing these references to our attention.}.

On the other hand, if $\Lambda<0$, then the 
new flat solution is stable. Can this 
be used at late times for the adjustment of the 
cosmological constant? One could be  contemplating  
a scenario in which a small region 
in a primordial universe first fluctuates into 
a state described by the new flat solution  
with a positive potential (positive $\Lambda$),   
then undergoes inflation
as described above,  and after that the potential drops 
to a negative  value $\Lambda<0$.  One  could use the new 
flat solution with $\Lambda<0$ to obtain an (almost) flat universe 
today via this sequence. We will briefly 
comment on what it takes to have such a scenario.

Before we turn to quantum cosmology of CGR, we would like to make a 
few comment concerning  the consistency of the theory 
(\ref {Lagr}) itself (this was discussed in detail in \cite {GShang},
here we just briefly summarize some main results):  

\begin{itemize} 

{\item  The Lagrangian (\ref {Lagr}) is not reparametrization 
invariant -- the new term completely restricts the symmetry. 
Nevertheless, the equivalence principle is preserved. 
The gauge condition (\ref {gauge}) allows {\it local},
point-dependent gauge transformations, that can be used to 
eliminate a nontrivial metric and connection  
in an infinitesimal neighborhood of any space-time point.}

{\item The linearized theory has  two propagating 
physical polarizations of a 
graviton. No negative-norm states or tachyons appear
in the quadratic action.}

{\item Bianchi identities enforce an additional condition on the 
Lagrange multiplier:  
$g^{\mu\nu}\partial_\mu\partial_\nu\lambda_\alpha=0$.
The latter has to be respected by all  solutions of the theory.}

{\item  Conventional solutions of GR (the Schwarzschild solution etc.,)
are also solutions of CGR. This is because the above solutions can be 
transformed  to a gauge where (\ref {gauge}) is fulfilled, 
and putting $\lambda_\mu=0$,  eq. (\ref {eq1}) is also satisfied.}

{\item  The structure of the Lagrangian (\ref {Lagr}) is not ruined
by quantum loop corrections since for small fluctuations on a given 
background it can be completed to a 
BRST invariant form introducing the Faddeev-Popov ghost. The latter
do not affect the classical solutions that we discuss.}


\end{itemize}

\section{Minisuperspace for  constrained gravity}

Computations in quantum cosmology are primarily  
performed in a minisuperspace approximation (for a review see, e.g., 
\cite {Vilenkin1}). In this section we develop a minisuperspace approach to
CGR. The metric for a spatially-flat universe in this 
approach takes the form:
\beq
ds^2 = N^2(t) dt^2 -a^2(t)\delta_{ij}dx^i dx^j\,.
\label{eq:mini_metric}
\eeq
Here both $N$ and $a$ are functions of $t$ only, and $i,j=1,2,3$.
  One difference from the conventional approach is 
that we will be working with general $N$ not necessarily equal to the 
unity. We will show  that $N$  is determined by $a$ because of   
the  constraint.

The corresponding Lagrangian density (\ref {Lagr}) takes the form:
\begin{equation}
\mathcal{L}=-a^3N\left[6\frac{1}{N^2}\left(\frac{\dot a}{a}\right)^2
+2\Lambda\right]
-2\left(\frac{a^3}{N}\right)\dot\lambda_0
+2aN\nabla\cdot\vec\lambda,
\end{equation}
where $\vec\lambda=(\lambda_1, \lambda_2,\lambda_3)$, and 
$\nabla\cdot\vec\lambda \equiv \delta^{ij}\partial_i \lambda_j.$ 
As a part of the rules of the minisuperspace reduction  
we require that $\lambda_i$'s are time independent 
functions of spatial coordinates $x_i$ only.  This rule is 
justified by the complete Hamiltonian description of the theory 
(\ref {Lagr}) which is  given in the appendix \ref{app:hamiltonian}.
One can check that solutions to such a theory only exist
when $\lambda_0$ is a function of the time coordinate 
$t$ alone,  and  $\nabla\cdot\vec\lambda$ is a space-time constant.

It is straightforward to find  the Hamiltonian density.  
For the canonical momentum conjugate to $a$, 
we obtain $\pi_a=-12 a\dot a/N$.   Moreover,  
\begin{gather}
\label{eq:constraint_0}
\pi_{\lambda_0}=-\frac{2a^3}{N},\\
\label{eq:constraint_N}
\pi_{N}=0\,.
\end{gather}
The above relations 
represent two primary constraints of the Hamiltonian formalism. 
Note that we are not introducing a conjugate momentum  for $\vec\lambda$, 
since it is  assumed to be time independent, and, therefore non-dynamical.
A more rigorous treatment is given  in the appendix \ref{app:hamiltonian}.
The total Hamiltonian density takes the form:
\begin{equation}
\begin{split}
\mathcal{H}_{total}=-\frac{N}{24 a}\pi_a^2
+2a^3N\left(\Lambda-\frac{\nabla\cdot\vec\lambda}{a^2}\right)
+ \alpha  \left(\pi_{\lambda_0}+\frac{2a^3}{N}\right)
+\beta \pi_{N},
\end{split}
\end{equation}
where $\alpha$ and $\beta $ are Lagrange multipliers enforcing
the primary constraints. Due to the Hamiltonian equations of 
motion  $ \alpha =\dot\lambda_0$
and  $\beta = \dot N$. 
Requiring that  the time variation of the two primary 
constraints  vanishes,  we obtain the equations
of motion for the inexpressible velocities $\dot N $ and $\dot\lambda_0$
\begin{gather}
\frac{\ud}{\ud t}\left(\frac{a^3}{N}\right)=0,\\
\label{eq:velocity_0}
\dot{\lambda}_0=\frac{N^2}{2a^3}\left[-\frac{\pi_a^2}{24a}
+2a^3\left(\Lambda-\frac{\nabla\cdot\vec\lambda}{a^2}\right)\right].
\end{gather}
As it could be checked directly, no further constraints emerge. 
On the surface of the existing constraints we can 
simplify the  Hamiltonian density 
\begin{equation}
\label{eq:hamiltonian_mini}
\mathcal{H}=-\frac{N}{24 a}\pi_a^2
+2a^3N\left(\Lambda-\frac{\nabla\cdot\vec\lambda}{a^2}\right).
\end{equation}
Let us discuss classical solutions of such a theory first.  
From \eqref{eq:velocity_0} we find that  $a^3/N=b$, where $b$
is an arbitrary constant. As we discussed already 
$\nabla\cdot\vec\lambda\equiv3k$ is also a constant.  
From equation \eqref{eq:velocity_0} we find
\begin{equation}
\dot\lambda_0=\frac{N}{2a^3}\mathcal{H}.
\end{equation}
Since both $a^3/N$ and $\mathcal{H}$ itself commute with $\mathcal{H}$,
so does $\dot\lambda_0$. Hence, for $b\neq 0 $ we get that  
$\dot\lambda_0={\e} /2b$, which  is a 
constant if $\e$ is an eigenvalue 
(energy density) of $\mathcal{H}$.  

For  further convenience  we introduce the ``conformal time''
\begin{equation}
\eta=\int^t N(t')\ud t'.
\end{equation}
Then, the equations of motion can be expressed in
the following familiar form:
\beq
\left(\frac{a'}{a}\right)^2 + {k\over a^2} = {\Lambda\over 3} -
{b \e \over 6 a^6}\,, \\
\frac{a''}{a}={\Lambda\over 3} +\frac{b\e}{3a^6}\,,
\eeq
where $'\equiv\ud/\ud\eta=\ud/N\ud t$.  Interestingly, 
in these equations the quantity  $\nabla\cdot\vec\lambda \equiv 3k$ 
plays the role similar to  a three-dimensional spatial curvature of GR. 
Additional terms on the r.h.s. are also due to the $\lambda_\mu$ 
field. These terms act  as a fluid with the equation of state 
$\rho =p = {-b\e/2a^6}$.  Unlike other dynamical fields, there 
are no fluctuations of $\lambda_\mu$. 
  
We will consider the following three solutions of the equations of motion:

\begin{enumerate}

\item
$\e =0$, $k\ne0$, one finds a spatially flat inflating solution,
where the scale factor $a$, as a function of conformal time
$\eta$, is identical to that of a closed dS universe;

\item
$-b\e =2k=\Lambda$ and $a=1$, one finds a flat Minkowski space-time
in spite of the fact that  $\Lambda\ne0$. This is the new flat solution 
described in the previous section. We consider
two physically different cases: $\Lambda>0$ and $\Lambda<0$;

\item
$\e =k=0$, gives  a conventional, spatially-flat  inflating de Sitter 
space-time.

\end{enumerate}

Below  we will study physical consequences of these solutions
\footnote{Note that a negative sign of 
the product $b{\e}$ corresponds to a positive energy density of the fluid. 
In general, on certain solutions $b{\e}$ could also take a positive sign 
producing a negative energy density fluid. However, this should not be 
a concern since the $\lambda_\mu$ field,  that give rise to this 
fluid, is not dynamical and does not fluctuate.}.

\section{Wave-function and creation probability}

We now turn to  the quantum mechanics of  
the Hamiltonian density given by \eqref{eq:hamiltonian_mini}.  
To do so we  promote all the fields in 
\eqref{eq:hamiltonian_mini} to operators 
with the prescription   $\pi_a=-i \delta/\delta a$
and $\pi_{\lambda_0}=-i\delta/\delta\lambda_0$.
The Lagrangian is an integral of the density $\mathcal{L}$ over 
the entire space on each time slice. To make the integral converge,
we will be discussing  a three-dimensionally  
flat space with a finite-size spatial boundary.  
Then, the integral 
\begin{equation}
v=\int\ud^3 x,
\end{equation}
is finite, and $v$ denotes  a spatial ``comoving volume'' on each 
time slice and is a fixed number.  The physical  3-volume is 
$v_p(t)=\int\sqrt{\gamma}\ud^3x=a^3(t) v$.
So far we have ignored the factor  $v\mpl^2$ in the action.
Restoring this  factor, the  Hamiltonian \eqref{eq:hamiltonian_mini} reads
\begin{equation}
\label{eq:hamiltonian_mini_v}
H=-\frac{N}{24 v\mpl^2 a}\pi_a^2
+2v \mpl^2 a^3N\left(\Lambda-\frac{\nabla\cdot\vec\lambda}{a^2}\right).
\end{equation}
Let us now suppose that $\qket\psi$ is an eigenstate of $H$ with an energy 
eigenvalue $E$.  
As we discussed in the previous section,  both $\pi_{\lambda_0}\sim 
a^3/N$  and $\nabla\cdot\vec\lambda$ commute with $H$.  Therefore, 
one can  always choose $\qket\psi$ to be an eigenstate of the 
above two operators with eigenvalues $b$ and $3k$ respectively.  
On such a state, one can replace the operator $N$ 
by $a^3/b$, and $\nabla\cdot\vec\lambda$ by $3k$.
Therefore,  most generically,  we are looking for states
\begin{equation}
\qket\psi=\int \ud a' \psi(a') \qket{a'}_a\otimes\qket{a'^3/b}_N
\otimes\qket{3k}_{\nabla\cdot\vec\lambda},
\end{equation}
where $\qket{a'}_a$ represents the eigenstate of the operator
$a$ with eigenvalue $a'$, and etc..   The ``wave-function'' $\psi$
is determined by the Wheeler-De Witt equation
\begin{equation}
\left[- \frac{\ud^2}{\ud a^2}
+2 A \left(3k a^2-\Lambda a^4\right)+\frac{ A bE}{a^2}
\right]\psi(a)=0, 
\end{equation}
where $A\equiv 24 \mpl^4 v^2$, and we have ignored the operator 
ordering  ambiguity.  The solution of this equation is equivalent 
to the wave-function of a particle with zero energy moving in
a one-dimensional potential.

\begin{figure}[!htb]
\begin{center}
\includegraphics[width=0.4\textwidth]{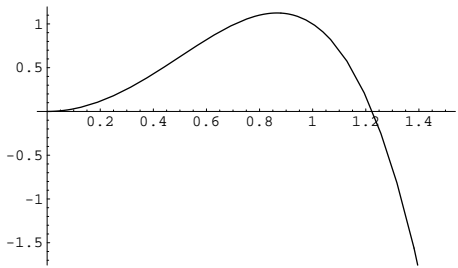}
\caption{\label{fig:zero} The potential $U(a)$.}
\end{center}
\end{figure}

Let us look now at a probability of  creation of the universe ``from 
nothing'' , i.e., from a state with no classical space-time \cite 
{Vilenkin}. The solution describing  this state
should have zero energy $E=0$.  In this case the minisuperspace   
potential $U(a)$ is shown in Fig. \ref{fig:zero}.  
It has a classically  forbidden region $0\le a\le \sqrt{3/\Lambda}$, 
and a  de-Sitter region $a\ge\sqrt{3/\Lambda}$.  

To calculate the 
probability of tunneling of the system from $a=0$ to $a=\sqrt{3/\Lambda}$
we follow Vilenkin's tunneling wave-function approach (first reference 
in \cite {Vilenkin} and \cite {Vilenkin1}). Linde's approach \cite {Linde}, 
although conceptually different, gives the same answer in this case.
Taking the trace of equation (\ref {eq1}) we easily find 
the action on the tunneling solutions  
\begin{equation}
\mathcal{L}=-{\mpl^2 \over 2}
\sqrt{-g}(R+2\Lambda+2g^{\mu\nu}\partial_\mu\lambda_\nu)
=2 {\mpl^2}\Lambda \sqrt{g},
\end{equation}
and, introducing the euclidean time $\tau$, we find
\begin{equation}
S_E=- 2 v \mpl^2\Lambda\int_0^{\sqrt{3/\Lambda}} |N(\tau)a^3(\tau)| 
\ud \tau\,.
\end{equation}
Although this looks similar 
to the result in conventional quantum cosmology
for the action of a closed dS universe,  
there is an essential difference. The comoving 
volume $v$ is not fixed by the value of the cosmological
constant $\Lambda$.  
As a result, if we are to maximize the probability of tunneling
by creating a  smallest size universe, then 
$v$ is only to be bounded by the Plank scale.
However, there  is the following consideration to be 
taken into account. The fate of the universe 
after creation will depend on the boundary conditions
chosen.  For the universe created out of ``nothing'' 
we assume  simple  ones 
that the boundary surface has no tension, 
and that there is no  exterior space-time\footnote{This space 
can be ``glued'' to its own copy past the boundary.}.  
Moreover, we adjust the value of the $\lambda_j$
field on the boundary so that the space is complete
(this is possible because the $\lambda_\mu$ field
enters only linearly trough its first derivative in 
the Lagrangian (\ref {Lagr})). 
Such a universe,  to continue its inflationary expansion, 
should have a size bigger or equal to the scale of its dS 
horizon $\sqrt{3/\Lambda}$, otherwise it would collapse 
\cite {Sasaki,Berezin,Aurilia,Guth}.  
This puts a lower bound on the size of an 
acceptable initial universe $v_p\gsim 1/\Lambda^{3/2}$.
For $\Lambda\ll \mpl^2$ we get that $|S_E|\gg 1$ and 
the quasi-classical arguments are well applicable.
For a given value of $\Lambda$, the tunneling probability
can be calculated using a conserved ``Klein-Gordon'' 
current $j_a =i(\psi^+\partial_a \psi -\psi \partial_a \psi^+)/2$
\cite {Vilenkin}, and takes the form  
$ P_T\propto \exp(-2|S_E|)$. The latter  will be maximized by 
a smallest  acceptable value of $v_p\sim 
1/\Lambda^{3/2}$. This gives a  results similar to the 
probability of creation of a closed dS universe in the tunneling 
approach  $P_T\sim\exp(-3\pi^2\mpl^2/\Lambda)$
\cite {Linde,Vilenkin}. However,  as was emphasized above, 
in the present context  the created universe  has zero 
spatial curvature, while in the conventional 
approach only a closed dS universe can be materialized  
``from nothing''.

The subsequent evolution of the created universe is clear. It will 
inflate and redshift away the contribution of the $\lambda_j$ field that 
played the role of the spatial-curvature  during the creation.

Note that if we instead followed a naive  Euclidean continuation of the 
partition function, we would have obtained for the probability 
$ln P \propto {\mpl^2 \Lambda v_p}$. 
In such  a case,  largest values  of 
$\Lambda$  and $v_p$ would have been preferred. This would favor 
a creation of an inflationary universe  of a huge size. 
The above prescription is similar to the Hartle-Hawking (HH) 
approach \cite {HH} because of the euclidean continuation 
(it also somewhat differs  from the HH no-boundary proposal 
since our solution has a boundary). However, it is not clear whether 
the obtained result  has an interpretation of a probability 
of creation of a universe form ``nothing''.

\section{Inflation through flat space}

As it was discussed  in Ref. \cite {GShang}, by choosing 
$\partial_{\mu}\lambda_{\nu}+ \partial_{\nu}\lambda_{\mu} 
=-\Lambda g_{\mu\nu}$ we find a Minkowski solution 
even though  $\Lambda$ is nonzero. We will examine the 
properties of this solution more closely in the present section.
Some results of the present section are similar to
those of \cite {LindeFlat} obtained  for topologically 
nontrivial compact universes \cite {ZS}.

\begin{figure}[!htb]
\begin{center}
\includegraphics[width=0.4\textwidth]{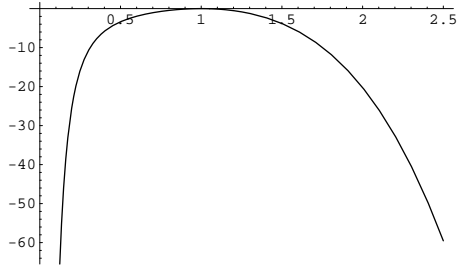}
\caption{\label{fig:mink} Potential $U(a)$ with an unstable 
space-time flat solution.}
\end{center}
\end{figure}

In the minisuperspace approach the above solution is described by 
$k=\Lambda /2$, $bE=- \Lambda$, $a=1$. Hence, the 
total energy of the solution is non-zero. 
Let us look at the mini-superspace potential  
$U$ with the above values of $k$ and $E$: 
\begin{equation}
U(a)=A\Lambda\left(3 a^2-2a^4-\frac{1}{a^2}\right).
\end{equation}
Here, as before,  $A=24\mpl^4 v^2$, and we put back 
$\mpl=1$ in this section.  This potential is  
illustrated in Fig. \ref{fig:mink}.  The new  Minkowski solution 
is described by the point  $a=1$. 
Such a universe cannot be created out of ``nothing'' since 
it has a nonzero energy. However, during some stage of 
the primordial evolution, for instance at the Planck scale,  
a part of space can fluctuate  into this state with an 
unsuppressed probability.  

What is the cosmological evolution of such a state?
Fluctuations in the system  will 
destabilize this state. It will either roll down  
toward $a=0$ corresponding to a contracting universe, 
or, with an almost equal probability, will roll  toward 
$a\rightarrow \infty$ corresponding to an inflating 
de Sitter  space.   It is easy to estimate the time scale of 
this instability.  Given any perturbation around $a=1$, the time 
scale it takes  for $a$ to change significantly is determined by 
$(\sqrt{|U''(1)|/12A\,})^{-1}=\sqrt{2/\Lambda}$. Therefore, the 
process of obtaining  
dS universe through the above flat solution, provides 
a new  channel for the  inflationary phase.

Let us now  discuss quantum mechanics of this model in more detail.
The Wheeler-De Witt equation reads 
\begin{equation}
\left[\frac{\ud^2}{\ud a^2}+A\Lambda
\left(2a^4-3 a^2+\frac{1}{a^2}\right)\right]\psi=0.
\end{equation}
When  $a\rightarrow +\infty$, the term $2 A \Lambda a^4$ 
dominates  over the other terms in the potential. 
Therefore,  the solution always asymptotes to a  
de Sitter universe and can be approximated by 
\begin{equation}
\psi\sim C_1 \sqrt a J_{+1/6}\left(\frac{\sqrt{2A\Lambda}}{3}a^3\right)
+C_2 \sqrt a J_{-1/6}\left(\frac{\sqrt{2A\Lambda}}{3}a^3\right),
\end{equation}
where $J_{\pm 1/6}$ are Bessel functions of the first kind.  The two
linearly independent solutions are  both oscillating and decaying.

The value of $A\Lambda$, however, can change the asymptotic 
behavior of $\psi$ in  the region of small $a$.  There are two 
possibilities.

\begin{enumerate}
\item
If we assume that $\Lambda\sim O(1)$    
and the volume of the universe at $a=1$ is $\sim O(1)$,  
we find $A\Lambda \sim 24$ (this is not a realistic case
since the absence of observed gravitational waves suggests 
that $H\sim \sqrt{\Lambda}$ has to be about 5 orders 
of magnitude below the Planck scale, nevertheless, we consider this 
as a theoretical example). Near this   region it's 
typical that $1-4A\Lambda<0$.  
In such a case the asymptotic behavior of $\psi$ near $a=0^+$ is 
given by 
\begin{equation}
\psi\sim C_1 \sqrt{a}\cos\left(\frac{\sqrt{4A\Lambda-1}}{2}\ln a\right)
+C_2 \sqrt{a}\sin\left(\frac{\sqrt{4A\Lambda-1}}{2}\ln a\right).
\end{equation}
These are oscillating solutions.  The  amplitude  scales 
as $\sqrt a$, and their frequencies increase to infinity toward the origin
at $a=0$.  Close to the origin the  wave-function 
has an  infinitely many zeros.
Since the amplitude of $\psi$ vanishes  
at the origin, this behavior should not be a concern.

The above two solutions differ only by a pure phase, and, therefore,
there is no physical reasons to favor one over the other.  With both
solutions allowed, one can smoothly  interpolate the 
wave-function and its first derivative from $a\sim 0^+$ to the 
region $a\sim +\infty$.  Typical solutions for  
$\psi$ in this case are shown in Fig. \ref{fig:unit_a}
and \ref{fig:unit_a_2}, with the $\cos$- and $\sin$-like initial conditions
near $a=0$ respectively.  The blue lines denote the  
potential $U(a)$.
\begin{figure}[!htb]
\begin{center}
\includegraphics[width=0.5\textwidth]{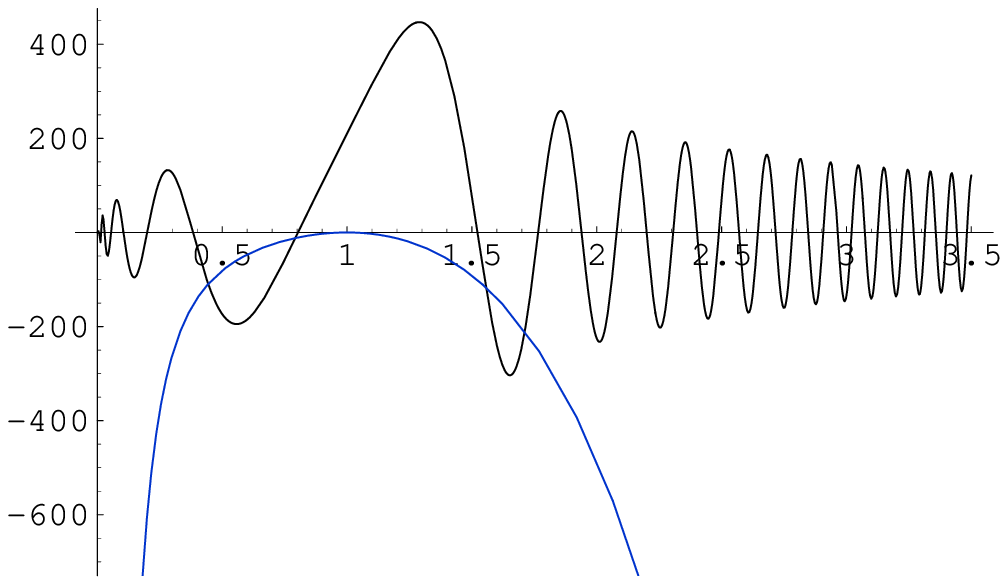}
\caption{\label{fig:unit_a} Solution with $A\Lambda=24$ and the $\cos$-like
initial condition}
\includegraphics[width=0.5\textwidth]{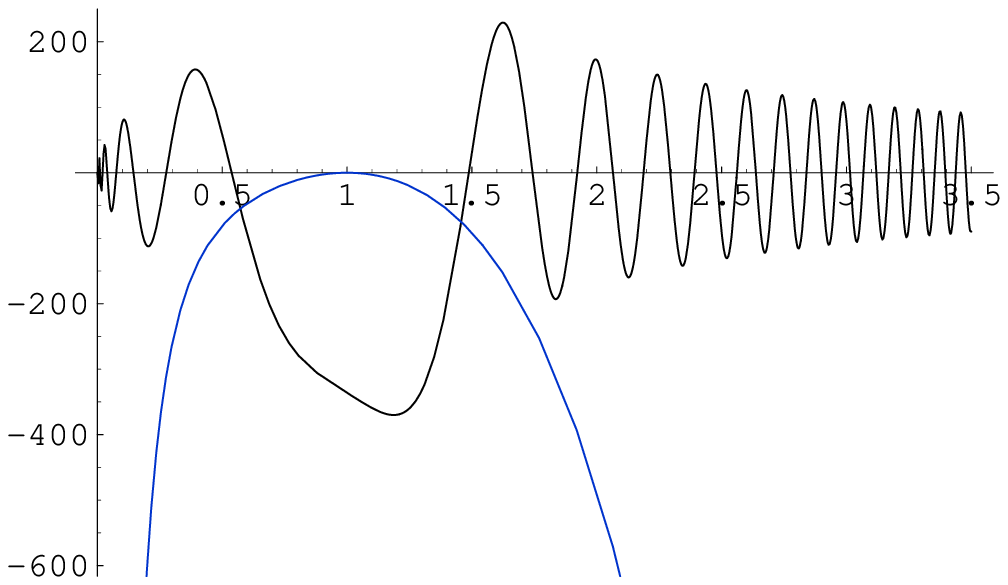}
\caption{\label{fig:unit_a_2} Solution with $A\Lambda=24$ and the $\sin$-like
initial condition}
\end{center}
\end{figure}

\item
If $\Lambda$ is much smaller, for example it is protected by 
supersymmetry at a scale much lower than the Planck scale, 
then the  product  $A\Lambda$ can be a very small number. In this 
case it is typical that $1-4A\Lambda>0$ and the asymptotic 
behavior of $\psi$ at $a=0^+$ changes to
\begin{equation}
\label{eq:small_a}
\psi\sim C_1 a^{\frac{1-\sqrt{1-4A\Lambda}}{2}}+
 C_2 a^{\frac{1+\sqrt{1-4A\Lambda}}{2}}.
\end{equation}
The wave-function  vanishes at $a=0$ since both exponentials are positive 
and perfectly regular.  Again, solutions that covers the entire 
region must exist since both solutions above are physically allowed. 

In the  limit where the quantity $A\Lambda$ is tiny one can 
ignore the existence of the potential $U(a)$ for a fairly long 
time, until $a$ grows and  $A\Lambda a^4$ becomes  comparable with 
$1$. When  this happens, $a$ is already so large  that all the  terms
in the potential, besides $2 A\Lambda a^4$, can be neglected.
The  wave function $\psi$ should quickly turn  into the Bessel functions 
described above.  Before that happens,  the Schr\"odinger
equation takes a simple form  $\psi(a)''=0$. 
As a result
\begin{equation}
\psi\sim C_1+C_2 a.
\end{equation}
In the present case, the contribution of the $1/a^2$ 
term in the potential is mostly ignorable except for the region very 
close to the origin.  It becomes important there only to fix the 
initial value of $\psi$.  Due to this term 
$\psi(0)$ can only be zero.

Typical properties of $\psi$ are show in Fig. \ref{fig:small_a} and
Fig. \ref{fig:small_a_2}.  One can see that the potential is
extremely flat until $a\gg 1$, after which it quickly takes a
form of $-2A\Lambda a^4$.  

\begin{figure}[!htb]
\begin{center}
\includegraphics[width=0.5\textwidth]{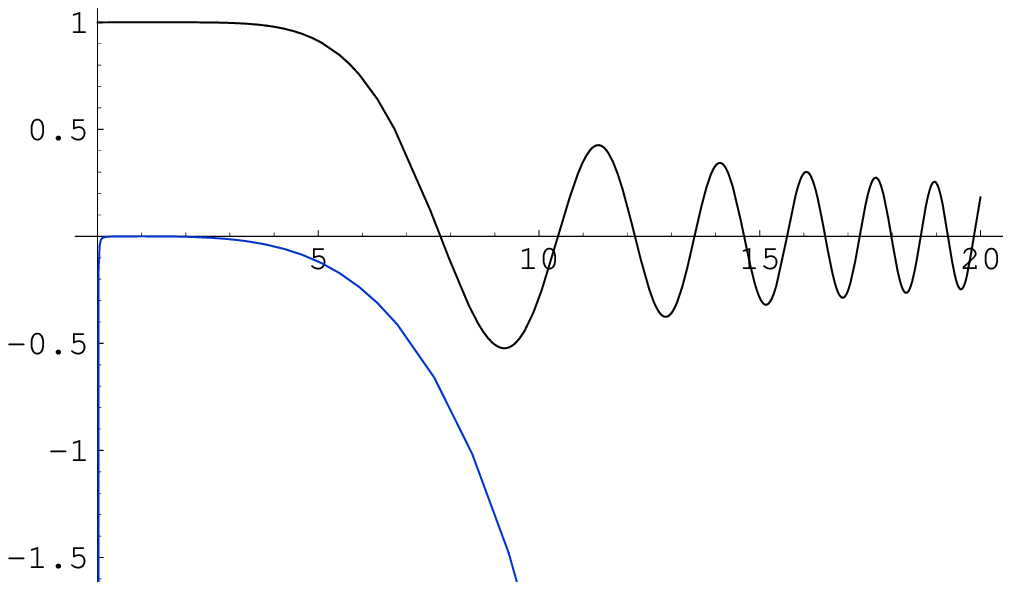}
\caption{\label{fig:small_a} Solution with $A=10^{-4}$;
$\psi$ tends to $0$ near $a=0$ too fast to be shown in this figure.}
\includegraphics[width=0.5\textwidth]{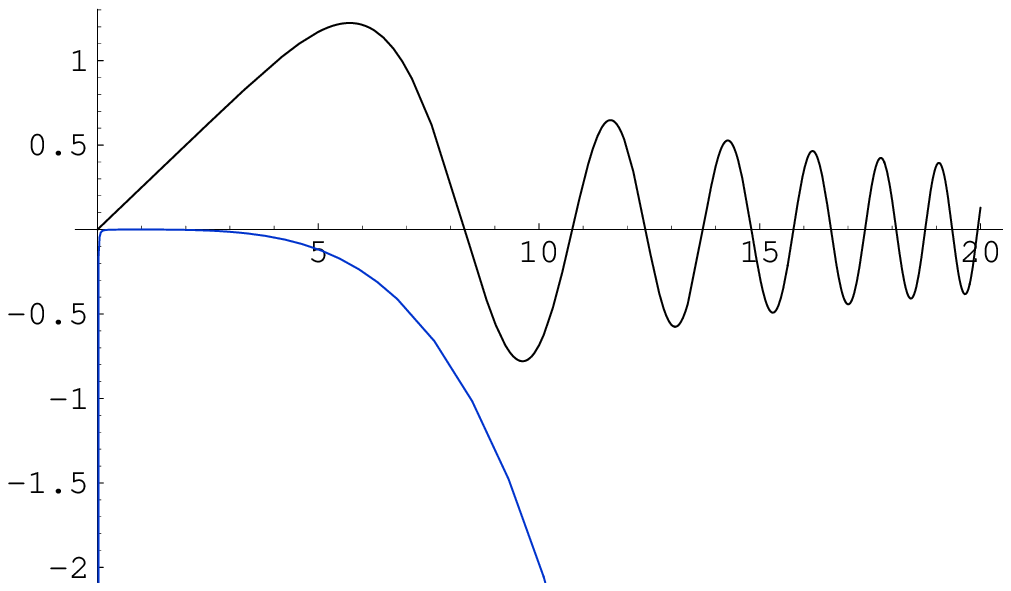}
\caption{\label{fig:small_a_2}Nearly linear solution with $A=10^{-4}$.}
\end{center}
\end{figure}

\end{enumerate}

\section{Conclusions}

Inflation provides a rather effective solution to the  problems of 
the hot big-bang cosmology and  successfully accounts for 
the observations (for a review and references, see 
\cite {LindeRev}). Under reasonable physical conditions, 
inflationary universe is not {\it past-eternal} \cite {Vilenkin2}, 
and one would like to specify the past boundary of an inflating region 
of space-time. Quantum cosmology is one framework in which 
this issue can be addressed (for a review, see \cite {Vilenkin1}). 
In this approach a {\it closed } dS universe can be materialized 
from ``nothing'', providing the initial conditions for inflation. 
The creation  probability for such a state is exponentially suppressed,
nevertheless, it favors inflationary initial conditions over
the conditions for a universe sitting at the 
bottom of the potential \cite {Linde}, \cite {Vilenkin}. 
On the other hand, if a {\it compact} spatially-flat dS universe 
of nontrivial topology is created \cite {ZS},\cite {LindeFlat}, the  
exponential suppression can go away. 
  
In the present work we discussed classically constrained gravity 
\cite {GShang}. This theory  arises upon path-integral 
quantization of gravity as a low-energy  field theory 
with certain boundary conditions (see discussions in Section 1 
and in Ref. \cite {GShang}). This approach  gives rise to new 
solutions of equations of motion,  some cosmological 
implications of which we studied in the present work.
We showed that a {\it spatially-flat} dS universe with a boundary 
can be created form ``nothing''. With simple    
boundary conditions that we choose, the probability
for creation of such a universe is exponentially suppressed, 
nevertheless, it favors inflationary initial conditions. This is 
similar to the result for a {\it closed } dS universe in the 
conventional approach  \cite {Linde,Vilenkin}. 

Furthermore, we found a new interesting channel in which the 
probability for the inflationary initial conditions is not 
exponentially suppressed.  The universe can fluctuate into a state 
with zero space-time curvature and then rapidly  transitions to 
the inflating  spatially-flat dS state.  The fact that 
the probability is not exponentially suppressed in this case 
is  similar to the finding of Ref. \cite {LindeFlat}, however, 
the context in which our results are obtained, 
and the details of the dynamics are different.

There are a few questions that we left out for future 
detailed studies.  It would be interesting to consider 
similar solutions in the presence of other dynamical fields,
scalars, fermions etc. For instance, if the cosmological constant in 
the Lagrangian (\ref {Lagr}) is  negative, then the new flat solution is 
stable.  One could imagine a scenario, in which  the original 
inflationary universe eventually ends up in a state with a negative 
value of the potential. In  that case, the  Lagrange multiplier field can 
neutralize the negative potential energy and gives rise to a stable flat
space-time. In general, however,  it is hard to maintain 
this state intact during the course of cosmological evolution since
any cosmological expansion of the universe redshifts the $\lambda$
terms rather quickly. Under these circumstances  
fine-tuning might be needed to obtain  a present-day  universe
in a state of the space-time flat solution with $\Lambda<0$.
The question of how severe this fine-tuning should be,  
and related issues will be discussed elsewhere.

\vspace{0.5in}
{\bf Acknowledgments}
\vspace{0.2in}

We would like to thank  Andrei Linde and Alexander Vilenkin
for valuable communications. The work was  supported in part by 
NASA grant NNG05GH34G, and in part by 
NSF grant PHY-0403005.  YS is also supported 
by graduate student  funds provided by New York University.  
YS would like  to thank  Andrei Gruzinov for his support 
through David and Lucile Packard Foundation Fellowship.

\vspace{0.5in}

\appendix
\section*{Appendices}

\section{Hamiltonian formalism for CGR}
\label{app:hamiltonian}
To find  the Hamiltonian we express the metrics in the  ADM formalism:
\begin{equation}
g_{\mu\nu}=\left(
\begin{array}{cc}
N^2-h_{ij} N^i N^j & -h_{ij} N^j \\
-h_{ji}N^i& -h_{ij}
\end{array}
\right), \quad
g^{\mu\nu}=\left(
\begin{array}{cc}
\frac{1}{N^2} & -\frac{N^i}{N^2} \\
-\frac{N^j}{N^2}& -h^{ij}+\frac{N^i N^j}{N^2}
\end{array}
\right).
\end{equation}
The Lagrangian density becomes (an overall factor of 2 is 
ignored below)
\begin{equation}
\begin{split}
\mathcal{L}=&\sqrt\gamma N(R^{(3)}+K_{ij}K^{ij}-K^2-2\Lambda)
-2\left[\left(\frac{\sqrt\gamma}{N}\right)\dot\lambda_0
-\left(\frac{\sqrt\gamma N^i}{N}\right)\partial_i\lambda_0\right] \\
&+2\left[\left(\frac{\sqrt\gamma N^i}{N}\right)\dot\lambda_i
+\left(N\sqrt\gamma h^{ij}-\frac{\sqrt\gamma N^i N^j}{N}\right)
\partial_j\lambda_i
\right],
\end{split}
\end{equation}
where ``$\cdot$''$\equiv\partial_0$. Here we have defined $\gamma=\det h_{ij}$ 
and the extrinsic curvature 
\begin{equation}
K_{ij}=\frac{1}{2N}(\dot h_{ij}-\uD_i N_j-\uD_j N_i).
\end{equation}
$\uD$ denotes the spatial covariant derivative defined w.r.t. $h_{ij}$.

In order to simplify the formalism we introduce the following new variables
\begin{equation}
\tilde N=\frac{\sqrt\gamma}{N},\qquad
\tilde N^i=\frac{\sqrt\gamma N^i}{N}.
\end{equation}
In terms of these new fields the Lagrangian density
reads
\begin{equation}
\begin{split}
\mathcal{L}=&\frac{\gamma}{\tilde N}(R^{(3)}+K_{ij}K^{ij}-K^2-2\Lambda)
-2(\tilde N\dot\lambda_0-\tilde N^i\partial_i\lambda_0)\\
&+2\left[\tilde N^i\dot\lambda_i
+\left(\frac{\gamma h^{ij}}{\tilde N}
-\frac{\tilde N^i\tilde N^j}{\tilde N}\right)\partial_j\lambda_i
\right].
\end{split}
\end{equation}
The conjugate momenta are:
\begin{gather}
\pi_{ij}=\sqrt\gamma (K_{ij}-K h_{ij}), \\
\label{eq:primary_constraints_0}
\pi_{\lambda_0}=-2\tilde N ,\\
\label{eq:primary_constraints_i}
\pi_{\lambda_i}=2\tilde N^i ,\\
\label{eq:primary_constraints_Ns}
\pi_{\tilde N}=\pi_{\tilde N^i}=0.
\end{gather}
Equations \eqref{eq:primary_constraints_0} 
through \eqref{eq:primary_constraints_Ns}
are to be understood as eight primary constraints 
to be imposed on physical states. 
The total Hamiltonian density, including all the inexpressible 
velocities, is given by
\begin{equation}
\begin{split}
\label{eq:hamiltonian_total}
\mathcal{H}_{total}=&-\frac{\gamma}{\tilde N} R^{(3)}
+\frac{1}{\tilde N}\left(\pi_{ij}\pi^{ij}-\frac{1}{2}\pi^2\right)
+2\pi^{ij}\uD_i\left(\frac{\tilde N_j}{\tilde N}\right)\\
&+\frac{2\gamma}{\tilde N}\Lambda-2\tilde N^i\partial_i \lambda_0
-2\left(\frac{\gamma h^{ij}}{\tilde N}-
\frac{\tilde N^i\tilde N^j}{\tilde N}\right)\partial_j\lambda_i\\
&+\pi_{\tilde N}\beta  
+\pi_{\tilde N^i}\gamma^i 
+(\pi_{\lambda_0}+2\tilde N) \alpha   
+(\pi_{\lambda_i}-2\tilde N^i)\delta_i\\
\equiv&\mathcal{H}_0+\pi_{\tilde N}\beta  
+\pi_{\tilde N^i}\gamma^i 
+(\pi_{\lambda_0}+2\tilde N) \alpha   
+(\pi_{\lambda_i}-2\tilde N^i)\delta_i.
\end{split}
\end{equation}
The definition of $\mathcal{H}_0$ can  easily be read off 
the expression above, and the Lagrange multipliers 
$\beta, \gamma_j, \alpha, \delta_i,$
are determined by the Hamilton equations 
in terms of the velocities 
$  \beta= \dot{\tilde N},~\gamma_j= \dot{\tilde N}_j,~
\alpha =\dot\lambda_0,~\delta_i= \dot\lambda_i.$
All the eight inexpressible velocities are resolvable:
\begin{gather}
\label{eq:velocity_N}
\dot{\tilde N}=\partial_i\tilde N^i,\\
\label{eq:velocity_N_i}
\dot{\tilde N}^i=-\partial_j\left(
\frac{\gamma h^{ij}}{\tilde N}-\frac{\tilde N^i\tilde N^j}{\tilde N}
\right),\\
\label{eq:velocity_lambda_0}
\dot\lambda_0=-\frac{1}{2}\frac{\delta\mathcal{H}_0}{\delta\tilde N},\\
\label{eq:velocity_lambda_i}
\dot\lambda_i=\frac{1}{2}\frac{\delta\mathcal{H}_0}{\delta\tilde N^i}.
\end{gather}
These results are very different from what one finds in GR. 
After eliminating the inexpressible velocities,  
the Hamiltonian density reads 
\begin{equation}
\label{eq:hamiltonian}
\begin{split}
\mathcal{H}=&
-\frac{\gamma}{\tilde N} R^{(3)}
+\frac{1}{\tilde N}\left(\pi_{ij}\pi^{ij}-\frac{1}{2}\pi^2\right)
+2\pi^{ij}\uD_i\left(\frac{\tilde N_j}{\tilde N}\right)\\
&+\frac{2\gamma}{\tilde N}\Lambda-2\tilde N^i\partial_i \lambda_0
-2\left(\frac{\gamma h^{ij}}{\tilde N}-
\frac{\tilde N^i\tilde N^j}{\tilde N}\right)\partial_j\lambda_i,
\end{split}
\end{equation}
where $\tilde N$ and $\tilde N^i$ \emph{must be 
identified with $-\pi_{\lambda_0}/2$ and $\pi_{\lambda_i}/2$ 
respectively while taking Poisson brackets}.  
We have thrown away terms that are 
proportional to $\pi_{\tilde N}$ and $\pi_{\tilde N^i}$, 
which necessarily vanish in any case.

Upon quantization one should impose the constraints $\pi_{\tilde N}
\qket{\psi}=\pi_{\tilde N^i}\qket{\psi}=0$ on physical
states and proceed with the usual canonical  procedure
using the Hamiltonian density given above.  Notice that since
$-2\tilde N\equiv\pi_{\lambda_0}$ and $2\tilde N^i\equiv\pi_{\lambda_i}$, 
and they both appear in $\mathcal{H}$, $\lambda_0$ and $\lambda_i$ do 
not commute with the Hamiltonian  and therefore are not conserved in general.
However, equation \eqref{eq:velocity_lambda_i} can be expressed as
\begin{equation}
\dot\lambda_i=-\frac{1}{\tilde N}\uD_j\pi^j_i
+\frac{\tilde N^j}{\tilde N}(\partial_j\lambda_i+\partial_i\lambda_j),
\end{equation}
therefore $\dot\lambda_i=0$ as long as $N^i=0$, and $\pi^i_j$
depends on time $t$ only.  In such a case, $\lambda_i$ do commute
with the Hamiltonian density.  It is because of this reason that 
we have imposed this condition in the minisuperspace formalism.

Mathematically such a constraint can be enforced  
more rigorously by introducing a term $A_i\dot{\lambda}_i$
in the Lagrangian density with Lagrange multipliers $A_i$.
Using this Lagrangian density one can work out $\mathcal{H}_{total}$ 
in a similar fashion as we did above.  After all the constraints 
are taken into account consistently, one finds that 
the Hamiltonian density is exactly the same as \eqref{eq:hamiltonian_mini}
with extra constraints $\pi_{A_i}=0$.  In the
quantum mechanics of such a theory one must then impose the
constraints $\pi_{A_i}\qket{\psi}=0$ on physical states $\qket\psi$.
This says that any physical state must be independent to the Lagrange 
multipliers $A_i$, which is  what one should have expected.

\section{Equations for $\lambda_\mu$ from Hamiltonian}
\label{app:constraints}
In the Hamiltonian formalism for CGR, time derivatives 
of some of  the primary constraints give rise to the equations
 \eqref{eq:velocity_N} and \eqref{eq:velocity_N_i}.
We will be using these conditions  in the following derivations
without mentioning them explicitly.  Time derivatives of the rest
of the  primary
constraints generate the equations of motion for the Lagrange multipliers
$\lambda_\mu$ (up to a surface term) as we will illustrate in this appendix.

From  \eqref{eq:velocity_lambda_0} and 
\eqref{eq:velocity_lambda_i} we have
\begin{gather}
\label{eq:eom_tN}
\dot\lambda_0=-\frac{1}{2}\frac{\partial \mathcal H_0}{\partial \tilde N}
=\frac{\mathcal H_0+2\tilde N^i\partial_i\lambda_0}{2\tilde N},\\
\label{eq:eom_tNi}
\dot \lambda_i=\frac{\partial\mathcal H_0}{2\partial \tilde N^i}
=-\frac{\sqrt\gamma}{\tilde N}\uD_j[(\sqrt\gamma)^{-1}\pi^j_i]
-\partial_i\lambda_0+\frac{\tilde N^j}{\tilde N}
(\partial_j\lambda_i+\partial_i\lambda_j).
\end{gather}
From the first equation above one immediately finds  that up to
a surface term
\begin{equation}
\mathcal{H}_0=2(\tilde N\dot\lambda_0+\lambda_0\dot{\tilde N})
=2\partial_0(\tilde N\lambda_0).
\end{equation}
Notice that $\partial_0\mathcal{H}_0=0$, and  
further time derivative of this equation gives
\begin{equation}
\begin{split}
0=&\int \ud^3x [\tilde N \ddot\lambda_0+2\dot{\tilde N}\dot\lambda_0
+\lambda_0\partial_i\dot{\tilde N}^i]\\
=&\int\ud^3x \left[\tilde N\ddot \lambda_0-2\tilde N^i\partial_i 
\dot\lambda_0-\left(\frac{\gamma h^{ij}}{\tilde N}
-\frac{\tilde N^i \tilde N^j}{\tilde N}\right)\partial_i\partial_j 
\lambda_0\right],
\end{split}
\end{equation}
which, up to a surface term, reproduces 
$g^{\mu\nu}\partial_\mu\partial_\nu\lambda_0=0$.

To make further use of equation \eqref{eq:eom_tNi} we first notice that
up to a surface term
\begin{equation}
2\int\ud^3x \sqrt\gamma\uD_j[(\sqrt\gamma)^{-1}\pi^j_i]
=\int\ud^3x h_{jk}\partial_i\pi^{jk}.
\end{equation}
Therefore, using the identities $\pi_{\lambda_0}=-2\tilde N$ and
$\pi_{\lambda_i}=2\tilde N^i$, the spatial integral of 
\eqref{eq:eom_tNi} can be simplified as 
\begin{equation}
\label{eq:eom_tNi_int}
-2\int\ud^3x [\tilde N\dot\lambda_i-\tilde N^j\partial_j\lambda_i]
=\int\ud^3x [h_{jk}\partial_i\pi^{jk}+\lambda_0\partial_i\pi_{\lambda_0}
+\lambda_j\partial_i\pi_{\lambda_j}].
\end{equation}
Likewise, we find that the time derivative 
of the l.h.s. of this equation gives
\begin{equation}
-2\int\ud^3x \left[\tilde N\ddot \lambda_i-2\tilde N^j\partial_j 
\dot\lambda_i-\left(\frac{\gamma h^{jk}}{\tilde N}
-\frac{\tilde N^j \tilde N^k}{\tilde N}\right)\partial_j\partial_k 
\lambda_i\right].
\end{equation}
To compute the time derivative of the r.h.s.  
of equation \eqref{eq:eom_tNi_int}
one only needs to notice that (up to a surface term)
\begin{equation}
\int\ud^3x[\dot h_{jk}\partial_i\pi^{jk}-\partial_i h_{jk}\dot\pi^{jk}]
=\int\ud^3x\left[\frac{\partial\mathcal H}{\partial \pi^{jk}}\partial_i\pi^{jk}
+\frac{\partial\mathcal H}{\partial h_{jk}}\partial_i h_{jk}\right].
\end{equation}
If we apply this same trick to the last two terms on the r.h.s. 
of equation 
\eqref{eq:eom_tNi_int} we find that its time derivative is simply
\begin{equation}
\begin{split}
\int\ud^3x&\left[\frac{\partial\mathcal H}{\partial \pi^{jk}}\partial_i\pi^{jk}
+\frac{\partial\mathcal H}{\partial h_{jk}}\partial_i h_{jk}
+\frac{\partial\mathcal H}{\partial \pi_{\lambda_0}}\partial_i\pi_{\lambda_0}
+\frac{\partial\mathcal H}{\partial \lambda_0}\partial_i \lambda_0 \right.\\
&\left.
+\frac{\partial\mathcal H}{\partial \pi_{\lambda_j}}\partial_i\pi_{\lambda_j}
+\frac{\partial\mathcal H}{\partial \lambda_j}\partial_i \lambda_j\right]\\
=&\int\ud^3x\partial_i\mathcal H.
\end{split}
\end{equation}
Here $\mathcal{H}$ is the Hamiltonian density given by \eqref{eq:hamiltonian}
in which $\tilde N$ and $\tilde N^i$ are understood as conjugate
momenta to $\lambda_0$ and $\lambda_i$ respectively.
Therefore, we find 
\begin{equation}
\tilde N\ddot \lambda_i-2\tilde N^j\partial_j 
\dot\lambda_i-\left(\frac{\gamma h^{jk}}{\tilde N}
-\frac{\tilde N^j \tilde N^k}{\tilde N}\right)\partial_j\partial_k 
\lambda_i=0,
\end{equation}
which is indeed equivalent to 
$g^{\mu\nu}\partial_\mu\partial_\nu\lambda_i=0$.

\end{document}